\pdfoutput=1

\documentclass[twocolumn,showpacs]{revtex4}
\usepackage{graphicx,amssymb,amsmath}
\usepackage[usenames,dvipsnames]{color}    
\begin{document}

\title{
  Minimal conductivity and signatures of quantum criticality in~ballistic graphene bilayer
}

\author{Grzegorz Rut}
\affiliation{Marian Smoluchowski Institute of Physics, 
Jagiellonian University, Reymonta 4, PL--30059 Krak\'{o}w, Poland}
\author{Adam Rycerz}
\affiliation{Marian Smoluchowski Institute of Physics, 
Jagiellonian University, Reymonta 4, PL--30059 Krak\'{o}w, Poland}

\begin{abstract}
We study the ballistic conductivity of graphene bilayer in the presence of next-nearest neighbor hoppings between the layers. An undoped and unbiased system was found in Ref.\ \cite{Rut14a} to show a~nonuniversal (length-dependent) conductivity $\sigma(L)$, approaching the value of $\sigma_\star=3/\pi\simeq{}0.95$ for large $L$. Here we demonstrate one-parameter scaling and determine the scaling function $\beta(\sigma)=d\ln{}\!\sigma/d\ln{}\!L$. The scaling flow has an attractive fixed point [$\,\beta(\sigma_\star)=0$, $\beta'(\sigma_\star)<0\,$] reproducing the scenario predicted for random impurity scattering of Dirac fermions with Coulomb repulsion, albeit the system considered is perfectly ballistic and interactions are not taken into account. The role of electrostatic bias between the layers is also briefly discussed.
\end{abstract}

\date{\today}
\maketitle

\section{Introduction} 
One of the most unexpected properties of graphene---a~two-dimensional form of carbon discovered in 2005 \cite{Nov05}---is the pseudodiffusive nature of charge transport via undoped ballistic samples, manifesting itself by the fact that dc conductance obeys the Ohm's law for classical conductors characterized by the universal quantum value of the conductivity \cite{Kat06a,Two06,Mia07,Dan08}, namely
\begin{equation}
\label{sigma0def}
  G=\sigma_0\,\frac{W}{L},\ \ \ \ \ \ \sigma_0=\frac{1}{\pi}
  \ \ \ \left[\frac{se^2}{h}\right]
\end{equation}
where $W$ is the sample width, $L$ is the length \cite{Ryc09}, and the units of $se^2/h$ with $s=4$ are chosen to account the spin and valley degeneracies. Such a~macroscopic quantum phenomenon has a~remarkable high-frequency analog, i.e., the visible light opacity of graphene also takes quantized values \cite{Nai08}. Although the opacity directly scales with the number of graphene layers, such an additive property usually does not apply for dc conductance \cite{Mac13}.

Early theoretical works on ballistic graphene bilayers \cite{Sny07,Cse07} showed that the minimal conductivity at zero bias situation changes abruptly as a~function of next-nearest neighbor interlayer hopping integral $t'$, taking the value of $\sigma_0=1/\pi$ for $t'=0$, or $\sigma_\star=3\sigma_0$ for any $t'\neq{}0$ \cite{Mog09}, provided that $s=8$ in Eq.\ (\ref{sigma0def}) due to the additional layer degeneracy. Appearance of such a~quantum critical behavior was attributed to the topological transition of the Fermi surface at low energies \cite{Mac13}. Experimental values of the minimal conductivity are generally lower than $\sigma_\star$, covering the range from $\sim\sigma_0$ \cite{Fel09} up to $2.5\,\sigma_0$ \cite{May11}. We have recently shown, employing the Landauer-B\"{u}ttiker formalism, that the minimal conductivity of finite, ballistic samples is not universal but length-dependent \cite{Rut14a}, and can be rationalized, for large $L$, as 
\begin{equation}
  \label{sigmalen}
  \sigma(L)\simeq\overline{\sigma}(L)=\sigma_\star
  \left[1-\left(\lambda/L\right)^\gamma\right],
  % \left[1-\left(\frac{\lambda}{L}\right)^\gamma\right],
\end{equation}
where the characteristic length $\lambda=\lambda(t')$, and  $\gamma<1$ is the parameter-independent exponent. In turn, the predictions of Refs.\ \cite{Sny07,Cse07} are restored for $L\rightarrow\infty$, whereas in the opposite limit ($L\rightarrow{}0$) one gets $\sigma(L)\rightarrow\sigma_0$ regardless $t'=0$ or $t'\neq{}0$. It is also shown in Ref.\ \cite{Rut14a} that the universal conductivity is restored for resonances with the Landau levels at high magnetic fields \cite{disofoo}. 

\begin{figure}[!h]
  \centerline{\includegraphics[width=0.9\linewidth]{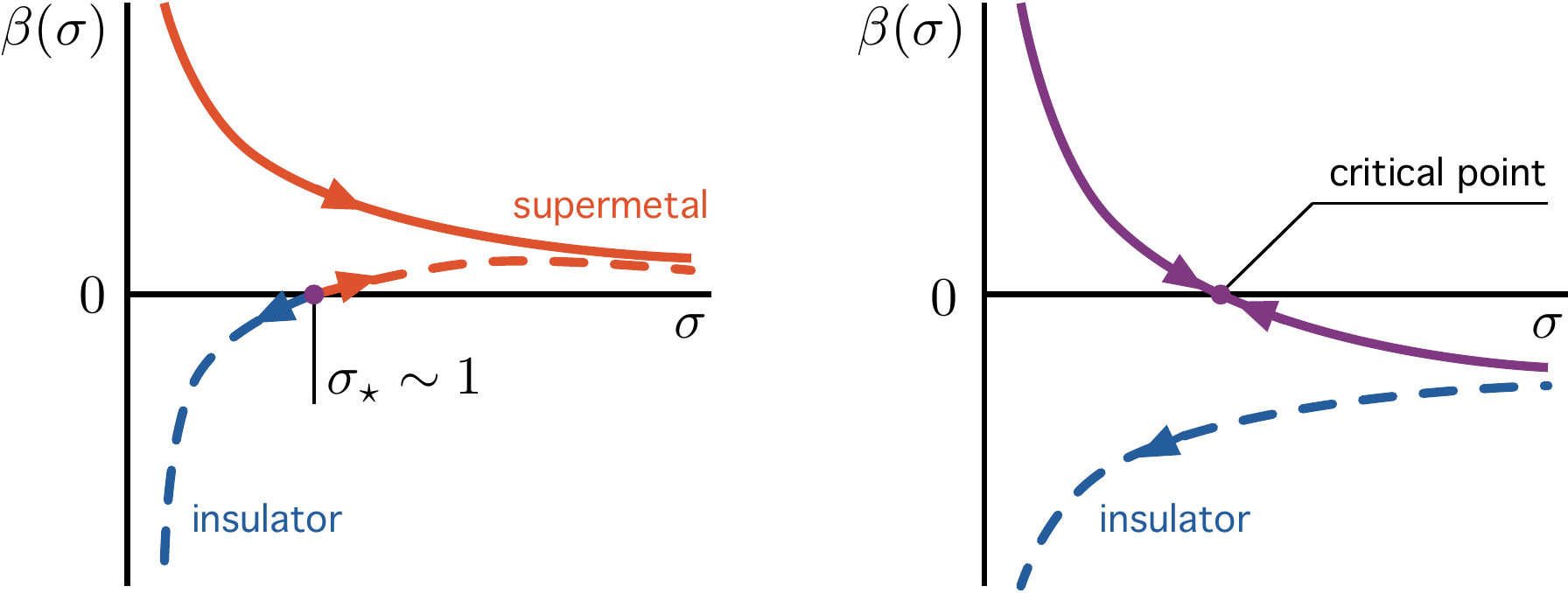}}
  \caption{ \label{betafigs}
    Schematic scaling functions $\beta(\sigma)$ (\ref{betadef}) for two-dimensional disordered Dirac (solid lines) and spin-orbit (dashed lined) systems.  Left: noninteracting case \cite{Bar07}, right: Coulomb interaction included \cite{Ost10}. Arrows indicate the flows of the dimensionless conductivity $\sigma$ with increasing $L$. (Adapted from Ref.\ \cite{Ost10}.)
}
\end{figure}

In this paper, we point out that the scaling function
\begin{equation}
  \label{betadef}
  \beta(\sigma)=\frac{d\ln\!\sigma}{d\ln\!L},
\end{equation}
which plays a~central role in conceptual understanding of the metal insulator transition \cite{Efe97} and is widely-considered in the context of disordered Dirac or spin-orbit systems \cite{Bar07,Eve08,Ost10} (see Fig.\ \ref{betafigs}), also unveils an intriguing analogy between interaction-induced quantum criticality in disordered Dirac systems \cite{Ost10} and transport properties of ballistic graphene bilayer with skew interlayer hoppings. 
The paper is organized as follows: 
In Section~II we present the mode-matching analysis for transport of Dirac fermions via finite samples of ballistic bilayer. In Section~III we discuss the functions $\sigma(L)$ for different values of $t'$ and demonstrate one-parameter scaling. Possible effects of nonzero bias between the layers are summarized in Section~IV. A~brief overview of the results given in Section~V. 

\begin{figure}[!ht]
  \centerline{\includegraphics[width=0.9\linewidth]{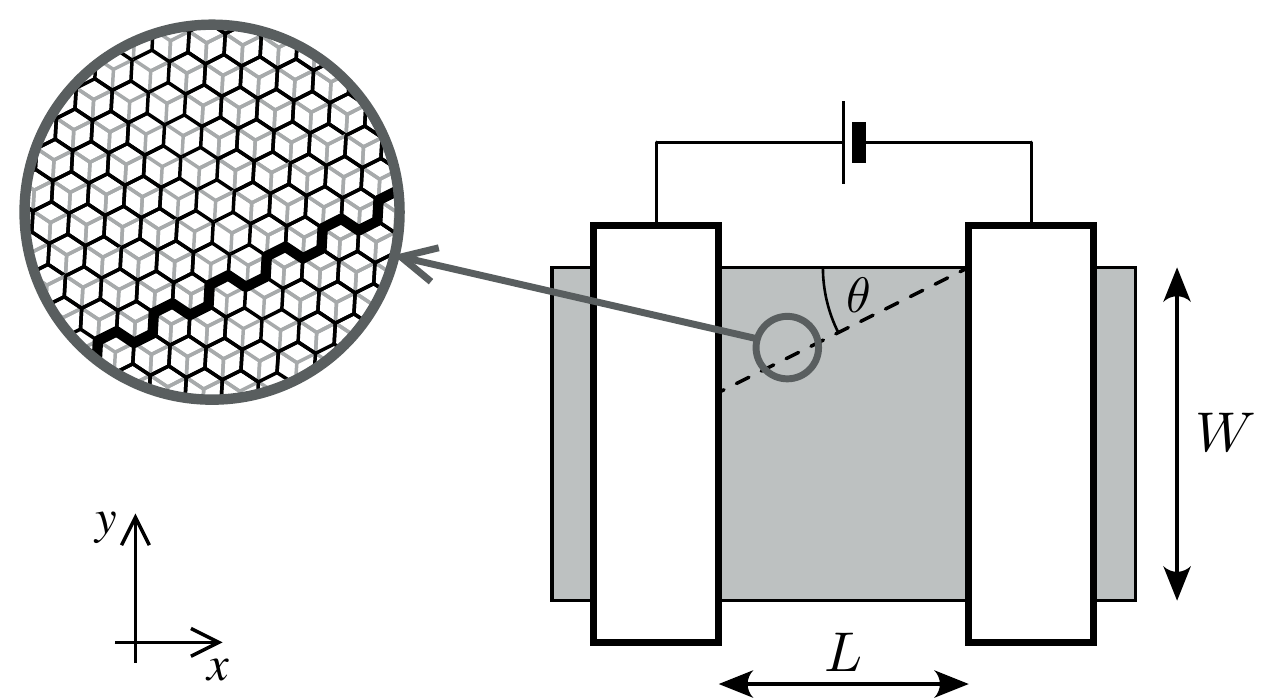}}
  \caption{  \label{setupfig}
    A~strip of bilayer graphene (shaded area) of width $W$, contacted by two electrodes (white rectangles) at a~distance $L$. A~voltage source drives electric current through the device. Separate top- and bottom-gate electrodes (not shown) allow both the Fermi energy $E$ and the bias between the layers $V$ to be controlled electrostatically. A~magnified view exhibits the crystallographic orientation, with the angle $\theta$ between an armchair direction (dashed line in the main plot) and $x$-axis of the coordinate system (see bottom left). 
  }
\end{figure}

%\section{Ballistic transport of Dirac fermions} 
\section{Mode-matching for Dirac fermions} 
The analysis starts from the four-band effective Hamiltonian for low-energy excitations \cite{Mac13}, which can be written as
\begin{equation}
  \label{ham1val}
  H=\xi\left(\begin{array}{cccc}
      -V/2 & e^{-i\theta}\pi & \xi{}t_{\perp} & 0\\
      e^{i\theta}\pi^{\dagger} & -V/2 & 0 & \nu{}e^{-i\theta}\pi\\
      \xi{}t_{\perp} & 0 & V/2 & e^{i\theta}\pi^{\dagger}\\
      0 & \nu{}e^{i\theta}\pi^{\dagger} & e^{-i\theta}\pi & V/2
    \end{array}\right),
%  H=\left(\begin{array}{cccc}
%      -V/2 & e^{-i\theta}\pi & t_{\perp} & 0\\
%      e^{i\theta}\pi^{\dagger} & -V/2 & 0 & \nu{}e^{-i\theta}\pi\\
%      t_{\perp} & 0 & V/2 & e^{i\theta}\pi^{\dagger}\\
%      0 & \nu{}e^{i\theta}\pi^{\dagger} & e^{-i\theta}\pi & V/2
%    \end{array}\right),
\end{equation}
where 
the valley index $\xi=1$ ($-1$) for $K$ ($K'$) valley, 
$\theta$ denotes the angle between the $x$-axis and the armchair direction (see Fig.\ \ref{setupfig}), $\pi=\hbar{}v_F(-i\partial_x+\partial_y)$ with $v_F=\sqrt{3}\,t_0a/2\simeq{}10^6\,$m/s being the Fermi velocity (in a~monolayer) defined via the interlayer hopping $t_0=3.16\,$eV  \cite{Kuz09} and the lattice parameter $a=0.246\,$nm, the nearest-neighbor interlayer hopping $t_\perp=0.38\,$eV, $\nu=t'/t_{0}$ with $t'$ being the next-nearest neighbor interlayer hopping, and $V$ is the electrostatic bias between the layers. We further consider solutions of the Dirac equation $H\Psi=E\Psi$ in the form 
\begin{equation}
  \label{psiansatz}
  \Psi(x,y)=\left(
    \begin{array}{c}
      \phi_1(x) \\ \phi_2(x) \\ \phi_3(x) \\ \phi_4(x)
    \end{array}
  \right)\exp(ik_{y}y)
\end{equation}
due to the translational invariance in the $y$-direction.

We focus here on a~zero bias case $V=0$ (for the discussion of $V\neq{}0$ case see Section~IV), for which a~general solution for $E=0$ (the sample area) and $K$ valley reads
\begin{widetext}
  \begin{equation}
    \label{phisample}
    \left(\begin{array}{c}
        \phi_{1}(x)\\
        \phi_{2}(x)\\
        \phi_{3}(x)\\
        \phi_{4}(x)
      \end{array}\right)=c_{1}\left(\begin{array}{c}
        -\alpha_{1}^{+}f_{1}^{+}\\
        0\\
        0\\
        e^{i\theta}f_{1}^{+}
      \end{array}\right)+c_{2}\left(\begin{array}{c}
        -\alpha_{-1}^{+}f_{-1}^{+}\\
        0\\
        0\\
        e^{i\theta}f_{-1}^{+}
      \end{array}\right)+c_{3}\left(\begin{array}{c}
        0\\
        e^{-i\theta}f_{-1}^{-}\\
        -\alpha_{-1}^{-}f_{-1}^{-}\\
        0
      \end{array}\right)+c_{4}\left(\begin{array}{c}
        0\\
        e^{-i\theta}f_{1}^{-}\\
        -\alpha_{1}^{-}f_{1}^{-}\\
        0
      \end{array}\right),
  \end{equation}
where 
$\alpha_{\zeta}^{\pm}=\nu\exp\left(\pm{}3i\theta\right)\left[i+\zeta\sqrt{\pm{}8ik_{y}/\left[\exp\left(\pm{}3i\theta\right)\tilde{t}\nu\right]+1}\right]$, 
$f_{\zeta}^{\pm}=\exp\left[\left(\pm k_{y}+\alpha_{\zeta}^{\pm}\tilde{t}\right)x\right]$,
$\tilde{t}=t_{\perp}/(\hbar v_{F})$,
and the coefficients $c_1,\dots,c_4$ are to be determined later. 
In the opposite limit of $E\rightarrow\infty$ (heavily-doped leads) we obtain 
\begin{equation}
  \label{phileads}
  \left(\begin{array}{c}
      \phi_{1,s}^{\pm}(x)\\
      \phi_{2,s}^{\pm}(x)\\
      \phi_{3,s}^{\pm}(x)\\
      \phi_{4,s}^{\pm}(x)
    \end{array}\right)={\cal N}_{\pm}(\nu)\exp\left(ik_{x}x\right)
  \left(\begin{array}{c}
      -\left(\mu_{\mp}/2\right)\exp\left(-2i\theta\right) \\
      s\eta_{\pm}\left(\mu_{\pm}/\sqrt{2}\right)\exp\left(-i\theta\right) \\
      s\sqrt{2}\eta_{\pm}\exp\left(i\theta\right)\\
      1
    \end{array}\right),
\end{equation}
where $s=\mbox{sgn}\left(k_{x}\right)$,  $\mu_{\pm}=\nu\pm\sqrt{\nu^{2}+4}$,  $\eta_{\pm}=1/\sqrt{2+\nu\mu_{\pm}}$, and the factors ${\cal N}_{\pm}(\nu)=\sqrt{\mu_{\pm}/\left[2\left(\mu_{\pm}+2\right)\right]}$ are chosen to normalized the current. Matching the solutions given by Eqs.\ (\ref{phisample},\ref{phileads}) at $x=0$ and $x=L$ leads to
\begin{equation}
\label{matchsys}
\begin{array}{c}
\left(\begin{array}{cccccccc}
\phi_{1,-1}^{+} & \phi_{1,-1}^{-} & \alpha_{1}^{+} & \alpha_{-1}^{+} & 0 & 0 & 0 & 0\\
\phi_{2,-1}^{+} & \phi_{2,-1}^{-} & 0 & 0 & -e^{-i\theta} & -e^{-i\theta} & 0 & 0\\
\phi_{3,-1}^{+} & \phi_{3,-1}^{-} & 0 & 0 & \alpha_{-1}^{-} & \alpha_{1}^{-} & 0 & 0\\
\phi_{4,-1}^{+} & \phi_{4,-1}^{-} & -e^{-i\theta} & -e^{-i\theta} & 0 & 0 & 0 & 0\\
0 & 0 & \alpha_{1}^{+}f_{1}^{+} & \alpha_{-1}^{+}f_{-1}^{+} & 0 & 0 & \phi_{1,1}^{+} & \phi_{1,1}^{-}\\
0 & 0 & 0 & 0 & -e^{-i\theta}f_{-1}^{-} & -e^{-i\theta}f_{1}^{-} & \phi_{2,1}^{+} & \phi_{2,1}^{-}\\
0 & 0 & 0 & 0 & \alpha_{-1}^{-}f_{-1}^{-} & \alpha_{1}^{-}f_{1}^{-} & \phi_{3,1}^{+} & \phi_{3,1}^{-}\\
0 & 0 & -e^{i\theta}f_{1}^{+} & -e^{i\theta}f_{-1}^{+} & 0 & 0 & \phi_{4,1}^{+} & \phi_{4,1}^{-}
\end{array}\right)\left(\begin{array}{c}
r_{p}^{\pm}\\
r_{n}^{\pm}\\
c_{1}^{\pm}\\
c_{2}^{\pm}\\
c_{3}^{\pm}\\
c_{4}^{\pm}\\
t_{p}^{\pm}\\
t_{n}^{\pm}
\end{array}\right)=\left(\begin{array}{c}
-\phi_{1,1}^{\pm}\\
-\phi_{2,1}^{\pm}\\
-\phi_{3,1}^{\pm}\\
-\phi_{4,1}^{\pm}\\
0\\
0\\
0\\
0
\end{array}\right)\end{array},
\end{equation}
\end{widetext}
where we have further defined $\phi_{j,s}^{\pm}\equiv\phi_{j,s}^{\pm}(0)$, $f_q^{\pm}=f_q^{\pm}(L)$. Solving the linear system of equations (\ref{matchsys}) one obtains the transmission and reflection matrices for a~given transverse wavenumber
\begin{equation}
\mathbf{t}(k_y)=\left(\begin{array}{cc}
      t_{p}^{+} & t_{n}^{+}\\
      t_{p}^{-} & t_{n}^{-}
    \end{array}\right), \ \ \ \ 
\mathbf{r}(k_y)=\left(\begin{array}{cc}
      r_{p}^{+} & r_{n}^{+}\\
      r_{p}^{-} & r_{n}^{-}
    \end{array}\right),
\end{equation}
where the internal structure arises from the presence of two subbands in the dispersion relation. At zero magnetic field, time-reversal symmetry coupling the valleys $K$ and $K'$ is preserved, and each transmission eigenvalue from one valley has a~copy in the other valley \cite{Bar08}. 

%\end{widetext}

\begin{figure}[!t]
\centerline{\includegraphics[width=0.9\linewidth]{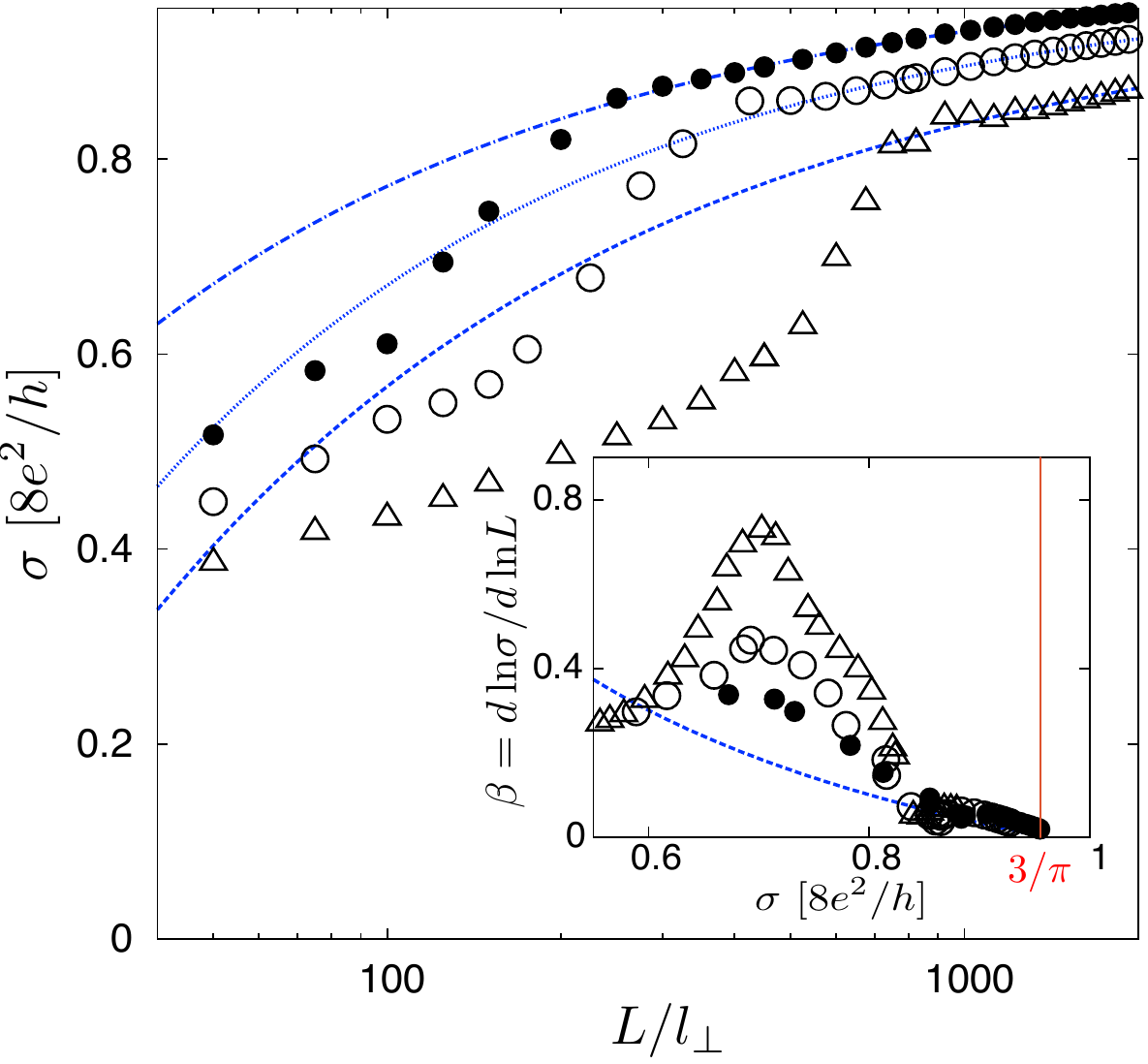}}
\caption{ \label{betasigm}
  Minimal conductivity of unbiased graphene bilayer as a~function of the sample length $L$ (specified in the units of $l_\perp=\hbar{}v_F/t_\perp\simeq{}1.60\,$nm). Different datapoints correspond to different values of the next-nearest neighbor interlayer hopping: $t'=0.1\,$eV ($\bigtriangleup$), $0.2\,$eV ($\bigcirc$), and $0.3\,$eV ({\Large\textbullet}). Lines depict best-fitted approximating functions $\overline{\sigma}(L)$ (\ref{sigmalen}) [see Table~I for further details]. {\em Inset} shows the scaling function $\beta(\sigma)$ (\ref{betadef}) obtained by numerical differentiation of the data. (Notice that the approximating function is replotted for $t'=0.1\,$eV only, as the three lines overlap for the variables used in the inset.)  The sample aspect ratio is fixed at $W/L=20$; the crystallographic orientation is $\theta=\pi/4$.
}
\end{figure}

\begin{table}
  \caption{
    Least-square fitted parameters $\sigma_\star$, $\lambda$, and $\gamma$ of the function $\overline{\sigma}(L)$, defined by Eq.\ (\ref{sigmalen}), corresponding to the lines in Fig.\ \ref{betasigm}. The values of $L_{0.01}$, such that for $L\geqslant{}L_{0.01}$ the function  $\overline{\sigma}(L)$ matches the actual conductivity with accuracy better that $1\%$, are given in the last column. 
  }
  \begin{tabular}{c|c|c|c|c}
    \hline\hline
    $\ t'\,[\mbox{eV}]\,$ & $\ \sigma_\star\,[8e^2\!/h]\,$ & $\ \lambda/l_\perp\,$
    & $\gamma$ & $\ L_{0.01}/l_\perp\,$ \\ \hline
    %%%  APPROXIMATED  %%%
    0.1 & $\ $0.96$\ $ & $\ $16.8$\ $ & $\ $0.50$\ $ & $\ $935$\ $  \\
    0.2 & 0.98 & 12.6 & 0.55 & 457 \\
    0.3 & 0.99 & $\ $6.1 & 0.53 & 278 \\    
    %%%  FULL PRECISION AS USED IN FIG.2  %%%
    % 0.1 & $\ $0.961$\ $ & $\ $16.83$\ $ & $\ $0.500$\ $  \\
    % 0.2 & 0.983 & 12.59 & 0.553 \\
    % 0.3 & 0.997 & $\ $6.08 & 0.532 \\
    \hline\hline
  \end{tabular} 
\end{table}

\section{Conductivity and one-parameter scaling} 
Next, the dimensionless conductivity is determined from the Landauer-B\"{u}ttiker formula \cite{Lan70} 
\begin{equation}
  \label{lansigma}
  \sigma(L)=\frac{L}{W}\sum_{\{k_y\}}\mbox{Tr}\left[
    \mathbf{t}(k_y)\,\mathbf{t}^\dagger\!(k_y)
  \right],
\end{equation}
where we have assumed periodic boundary conditions at $y$-direction, leading to the quantization of transverse wavenumber $k_y=0,\pm{}2\pi/W,\pm{}4\pi/W,\dots$. Our numerical results for $E=V=0$ are summarized in Fig.\ \ref{betasigm} and Table~I. For demonstrative purposes, we have chosen $W/L=20$ and $\theta=\pi/4$. For any $t'\neq{}0$, the conductivity given by Eq.\ (\ref{lansigma}) slowly grows with $L$, taking the values from the interval $1/\pi\leqslant{}\sigma(L)\leqslant{}3/\pi$, with the upper (lower) bound approached for $L\rightarrow{}0$ ($L\rightarrow{}\infty$). The best-fitted approximating functions $\overline{\sigma}(L)$ of the form given by Eq.\ (\ref{sigmalen})  [lines in Fig.\ \ref{betasigm}] rationalize the numerical results for large $L$ [datapoints]. The least-square fitted parameters $\sigma_\star$ and $\gamma$ [see Table~I] weakly depend on $t'$, taking the values close to $\sigma_\star\simeq{}3/\pi$ and $\gamma\simeq{}1/2$ for small $t'$. These scaling characteristics appear generically for other crystallographic orientations $\theta$, except from $\theta=\pi/6+n\pi/3$ (with integer $n$), corresponding to the propagation along a~zigzag direction, for which a~lower value of $\gamma\simeq{}1/4$ was found (see also Ref.\ \cite{Rut14a}).

The scaling function $\beta(\sigma)$ (\ref{betadef}) can be obtained by numerical differentiation of $\sigma(L)$ given by Eq.\ (\ref{lansigma}) [see inset in Fig.\ \ref{betasigm}]. For the asymptotic range, Eq.\ (\ref{sigmalen}) leads to 
\begin{equation}
  \label{betasim}
  \beta(\sigma)\simeq{}
  -\gamma\left(  1-\sigma_\star/\sigma  \right),
\end{equation}
with $\sigma=\sigma_\star\simeq{}3/\pi$ being an attractive fixed point [$\beta'(\sigma_\star)<0$ for $\gamma\simeq{}1/2>0$] of the renormalization group flow. Such a~scenario, earlier predicted for disordered Dirac systems with Coulomb interaction \cite{Ost10}, is reproduced by our results for graphene bilayer. The values of $\beta(\sigma)$ obtained numerically become $t'$-independent and follow Eq.\ (\ref{betasim}) for $\sigma\gtrsim{}0.8$. 

This surprising coincidence (it is worth to stress here that the system we consider is ballistic and no interactions are taken into account) seems difficult to understand in terms of existing symmetry-based theory of localization \cite{Eve08,Ost10}. Particular features of the results suggest that next-nearest neighbor interlayer hoppings, apart from breaking the rotational symmetry of the Hamiltonian in a~single valley (a~phenomenon know as {\em trigonal warping} \cite{Mac13}), may also induce corrections to $\beta(\sigma)$ of the Altshuler-Aronov type \cite{Alt85}, destroying the supermetalic phase in graphene. A~further clarification of the above-mentioned issue requires numerical study of charge transport through the disordered graphene bilayer, which is beyond the scope of this paper. 

% {\color{red} Moze dopisac '$1/2$' we wzorze (\ref{lansigma}) i~zmienic jednostki w~tabelach i rysunkach: $8e^2h\ \rightarrow\ 4e^2/h\ $?} 

\begin{figure}[!t]
\centerline{\includegraphics[width=\linewidth]{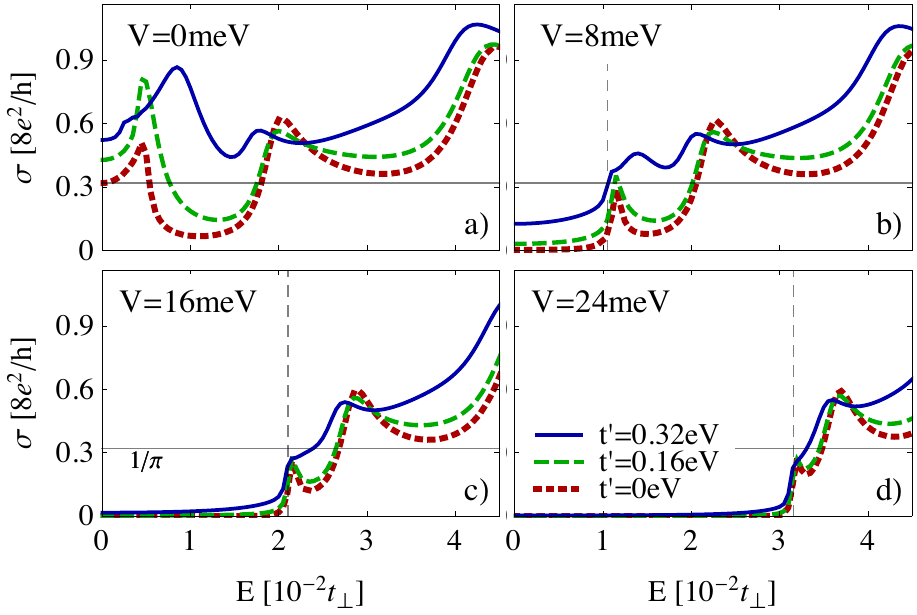}}
\caption{ \label{numgrapha}
  Conductivity of graphene bilayer as a~function of the Fermi energy. The value of bias between the layers $V$ is varied between the panels. Different lines at each panel depict the data obtained numerically for different values of the next-nearest neighbor interlayer hopping: $t'=0$ (red dashed line), $t'=0.16\,$eV (green dash-dot line), and $t'=0.32\,$eV (blue solid line). Vertical dashed line marks the value of $E=V/2$, horizontal solid line corresponds to $\sigma_0=1/\pi$. The sample length is fixed at $L=48\,l_\perp\simeq{}77\,$nm; the remaining system parameters are same as in Fig.\ \ref{betasigm}.
}
\end{figure}

\section{Effects of finite bias between the layers} 
Probably, the most intriguing property of graphene bilayer, is the possibility to convert it from semimetal to narrow-gap semiconductor by applying a~perpendicular electrostatic field \cite{Mac06,Oht06,Oos07,Cas07,Zha09}, leading to a~finite bias between the layers $V$ in effective Hamiltonian $H$ (\ref{ham1val}). Also, some experimental works showed that the energy gap may also appear spontaneously, due to electron-electron interactions, for bilayer samples close to the charge-neutrality point \cite{Rtt11,Bao12}. For these reasons, the extension of our discussion on the $V\neq{}0$ case is desirable.

In such a~case, the effective Dirac equation $H\Psi=E\Psi$, with $\Psi(x,y)$ in the form given by Eq.\ (\ref{psiansatz}), is integrated numerically for the sample area ($0<x<L$), separately for each value of the transverse wavenumber $k_y$. The obtained solutions are then matched with wavefunctions in the leads [see Eq.\ (\ref{phileads})], in analogy with the procedure presented in Section~II. 

The resulting conductivity spectra, for selected values of $V$ and $t'$, are displayed in Figs.\ \ref{numgrapha}(a)--\ref{numgrapha}(d).  For demonstrating purposes the sample dimensions are fixed at $W/L=20$, $L=48\,\hbar{}v_F/t_\perp$. For the unbiased sample case [see Fig.\ \ref{numgrapha}(a)] the conductivity systematically grows with the increasing $t'$ in the small vicinity of the Dirac point, a~width of which can be roughly approximated by $|E|\lesssim{}E_{\rm L}$, with 
\begin{equation}
  E_{\rm L}=\frac{1}{4}\,t_\perp\left(t'/t_0\right)^2
\end{equation}
being the Lifshitz energy \cite{Mac13}, reaching the value of $E_{\rm L}\simeq{}1\,$meV for $t'=0.32\,$eV. For $V=0$ and higher Fermi energies, $\sigma$ is weakly affected by $t'$. For $V>0$, the conductivity is strongly suppressed in the range of $|E|<V/2$ for any $t'$ [see Figs.\ \ref{numgrapha}(b)--\ref{numgrapha}(d)], provided that $V\gg{}E_{\rm L}$. Again, for sufficiently high energies the conductivity is almost unaffected by either the value of $V$ or $t'$. 

Probably, the most interesting feature of the results presented in Fig.\ \ref{numgrapha}, is that the dimensionless conductivity at its first maximum as a~function of $E$, reaches the value close to $\sigma\simeq{}1/\pi$ for $V\gg{}E_{\rm L}$ and arbitrary $t'$, while it is significantly higher for $V=0$. For this reason, the measurements of the conductivity spectra of ballistic samples at zero magnetic field, and different biases between the layers, may constitute an alternative experimental method for detecting the Lifshitz transition in graphene bilayer, supplementing the recent study focusing on the anomalies in the sequence of Landau levels \cite{Var14}, at least in principle.

%\section{Conclusions} 
\section{A brief overview}
We have investigated, by means of analytical mode-matching for the effective Dirac equation, the length-dependent minimal conductivity $\sigma(L)$ of {\em unbiased} graphene bilayer with the nearest ($t_\perp$) and the next-nearest neighbor ($t'$) interlayer hoppings included. The scaling function $\beta(\sigma)=d\ln{}\!\sigma/d\ln{}\!L$ was found (i) to be insensitive to the precise value of $t'$ and to the crystallographic orientation of the sample, provided that the physical dimensions are in the asymptotic range, i.e., that $W\gg{}L\gg{}\hbar{}v_F/t_\perp$ (with $v_F$ being the energy-independent Fermi velocity in a~monolayer), and (ii) to have an attractive fixed point at $\sigma_\star\simeq{}3/\pi$. These features closely resemble quantum critical behavior predicted theoretically for disordered Dirac systems with Coulomb interaction \cite{Ost10}, although the system we consider is ballistic and interactions are not taken into account. Our results show that the well-known correspondence between charge-transfer characteristics of classical diffusive conductor and perfectly clean monolayer graphene \cite{Two06,Ben08} is accompanied by another, probably more surprising, analogy between chaotic impurity scattering of interacting Dirac fermions and ballistic transport via bilayer samples, which gets unveiled when one-parameter scaling is demonstrated. 

The actual effects of electron-electron interaction in bilayer graphene are generally beyond the scope of this paper. Nevertheless, it is worth to point out that possible effects primarily include the gap opening due to spontaneous breaking of the symmetry between the layers \cite{Rtt11,Bao12}. We show, analysing numerically the transport through {\em biased} bilayer, that opening a~few meV gap $V$ (i.e., larger than the Lifshitz energy) leads to the appearance of conductivity peaks at Fermi energies $E\simeq{}\pm{}V/2$, where $\sigma\simeq{}1/\pi$, reproducing the dimensionless conductivity of a~ballistic monolayer. These are the reasons, for which an extensive experimental study of size-dependent conductance for clean bilayer samples with lengths $L>1\,\mu$m and widths $W\gg{}L$, which is missing so far, seems crucial to determine the significance of the factors such as the trigonal warping and the electron-electron interaction in effective description of bilayer-based graphene nanodevices \cite{gapfoo}.

\section*{Acknowledgments} 
The work was supported by the National Science Centre of Poland (NCN) via Grant No.\ N--N202--031440, and partly by Foundation for Polish Science (FNP) under the program TEAM {\em ``Correlations and coherence in quantum materials and structures (CCQM)''}. Computations were partly performed using the PL-Grid infrastructure.

%%%%%%%%%%%%%%%%%%%%%%%%%%%%%%%%%%%%%%%%%%%%%%%%%%%%%%%%%%%%%%%%%%%%%%%%%%%%%%
%%%%%%%%%%%%%%%%%%%%%%%%%%%%%%%%%%%%%%%%%%%%%%%%%%%%%%%%%%%%%%%%%%%%%%%%%%%%%%


\begin{thebibliography}{}

\bibitem{Rut14a}
G.~Rut and A.~Rycerz, Phys.\ Rev.\ B {\bf 89}, 045421 (2014). 

\bibitem{Nov05}
K.S.~Novoselov, A.K.~Geim, S.V.~Morozov {\em et al.}, Nature {\bf 438}, 197 (2005);  
Y.~Zhang, Y.-W.~Tan, H.L.~Stormer, and P.~Kim, {\it ibid.} {\bf 438}, 201 (2005).

\bibitem{Kat06a}
M.I.~Katsnelson, Eur.\ Phys.\ J.\ B \textbf{51}, 157 (2006).

\bibitem{Two06} 
J.~Tworzyd{\l}o, B.~Trauzettel, M.~Titov, A.~Rycerz, and C.W.J.~Beenakker, 
Phys.\ Rev.\ Lett.\ \textbf{96}, 246802 (2006).

\bibitem{Mia07} 
F.~Miao, S.~Wijeratne, Y.~Zhang, U.C.~Coscun, W.~Bao, and C.N.~Lau, 
Science \textbf{317}, 1530 (2007).

\bibitem{Dan08}
R.~Danneau, F.~Wu, M.F.~Craciun, S.~Russo {\em et al.}, 
Phys.\ Rev.\ Lett.\ \textbf{100}, 196802 (2008).

\bibitem{Ryc09}
The rectangular sample geometry is assumed for simplicity. 
For a~generalization, see:
A.~Rycerz, P.~Recher, and M.~Wimmer, Phys.\ Rev.\ B {\bf 80}, 125417 (2009).

\bibitem{Nai08}
R.R.~Nair, P.~Blake, A.N.~Grigorenko {\em et al.}, 
Science {\bf 320}, 1308 (2008); 
T.~Stauber, N.M.R.~Peres, and A.K.~Geim, 
Phys.\ Rev.\ B {\bf 78}, 085432 (2008).

\bibitem{Mac13}
E.~McCann and M.~Koshino, Rep.\ Prog.\ Phys.\ {\bf 76}, 056503 (2013).

\bibitem{Sny07}
I.~Snyman and C.W.J.~Beenakker, Phys.\ Rev.\ B \textbf{75}, 045322 (2007).

\bibitem{Cse07}
J.~Cserti, A.~Csord\'{a}s, and G.~D\'{a}vid, 
Phys.\ Rev.\ Lett.\ {\bf 99}, 066802 (2007); 
M.~Koshino and T.~Ando, Phys.\ Rev.\ B {\bf 73}, 245403 (2006).

\bibitem{Mog09}
Minimal conductivity $\sigma_\star=3\sigma_0$ appears for a~generic crystallographic orientation of a~sample. In a~particular case when the current is passed precisely along a~zigzag direction, the lower value of $(7/3)\,\sigma_0$ is predicted; see:
A.G.~Moghaddam and M.~Zareyan, Phys.\ Rev.\ B {\bf 79}, 073401 (2009).

\bibitem{Fel09}
B.~Feldman, J.~Martin, and A.~Yacoby, Nature Phys.\ {\bf 5}, 889 (2009).

\bibitem{May11}
A.S.~Mayorov, D.C.~Elias, M.~Mucha-Kruczy\'{n}ski {\em et al.},
Science {\bf 333}, 860 (2011).

\bibitem{disofoo}
Remarkably, some numerical studies of {\em disordered} monolayer samples reports the longitudinal conductivity close to $\sigma_0$, appearing at each Landau level for wide ranges of disorder and magnetic fields, see: 
F.~Ortmann and S.~Roche, Phys.\ Rev.\ Lett.\ {\bf 110}, 086602 (2013).

\bibitem{Efe97}
K.~Efetov, {\it Supersymmetry in Disorder and Chaos} (Cambridge University, Cambridge, 1997).

\bibitem{Bar07}
J.H.~Bardarson, J.~Tworzyd\l{}o, P.W.~Brouwer, and C.W.J.~Beenakker, 
Phys.\ Rev.\ Lett.\ {\bf 99}, 106801 (2007);
K.~Nomura, M.~Koshino, and S.~Ryu, 
{\em ibid.} {\bf 99}, 146806 (2007).

\bibitem{Eve08}
F.~Evers and A.D.~Mirlin, Rev.\ Mod.\ Phys.\ {\bf 80}, 1355 (2008).

\bibitem{Ost10}
P.M.~Ostrovsky, I.V.~Gornyi, and A.D.~Mirlin, 
Phys.\ Rev.\ Lett.\ {\bf 105}, 036803 (2010).

\bibitem{Kuz09}
Tight-binding parameters are taken from:
A.B.~Kuzmenko, I.~Crassee, D. van der Marel, P.~Blake and K.S.~Novoselov, 
Phys.\ Rev.\ B {\bf 80}, 165406 (2009).

\bibitem{Bar08}
J.H.~Bardarson, 
J.\ Phys.\ A: Math.\ Theor.\ {\bf 41}, 405203 (2008).

\bibitem{Lan70}
R.~Landauer, Phil.\ Mag.\ {\bf 21}, 863 (1970); 
M.~B\"{u}ttiker, Phys.\ Rev.\ B {\bf 46}, 12485 (1992).

\bibitem{Alt85}
B.L.~Altshuler and A.G.~Aronov, in {\it Electron-Electron Interactions in Disordered Conductors}, edited by A.L.~Efros and M.~Pollak (Elsevier, New York, 1985), p.\ 1.

\bibitem{Mac06}
%E.~McCann and V.I.~Fal'ko, Phys.\ Rev.\ Lett.\ {\bf 96}, 086805 (2006); 
E.~McCann, Phys.\ Rev.\ B {\bf 74}, 161403(R) (2006). 

\bibitem{Oht06}
T.~Ohta, A.~Bostwick, T.~Seyller, K.~Horn, and E.~Rotenberg, Science {\bf 313}, 951 (2006).

\bibitem{Oos07}
J.B.~Oostinga, H.B.~Heersche, X.~Liu, A.F.~Morpurgo, and L.M.K.~Vandersypen, Nat.\ Mater.\ {\bf 7}, 151 (2007). 

\bibitem{Cas07}
E.V.~Castro, K.S.~Novoselov, S.V.~Morozov, N.M.R.~Peres, J.M.B.~Lopes Dos Santos, J.~Nilsson, F.~Guinea, A.K.~Geim, and A.H.~Castro Neto, Phys.\ Rev.\ Lett.\ {\bf 99}, 216802 (2007).

\bibitem{Zha09}
Y.~Zhang, T.-T.~Tang, C.~Girit, Z.~Hao, M.C.~Martin, A.~Zettl, M.F.~Crommie, Y.R.~Shen, and F.~Wang, Nature {\bf 459}, 820 (2009).

\bibitem{Rtt11}
G.M.~Rutter, S.~Jung, N.N.~Klimov, D.B.~Newell,	N.B.~Zhitenev, and J.A.~Stroscio, Nature Physics {\bf 7}, 649 (2011). 

\bibitem{Bao12}
W.~Bao, J.~Velasco, Jr., F.~Zhang, L.~Jing, B.~Standley, D.~Smirnov, M.~Bockrath, A.H.~MacDonald, and C.N.~Lau, Proc.\ Natl.\ Acad.\ Sci.\ U.\ S.\ A.\ {\bf 109}, 10802 (2012). 

\bibitem{Var14}
A.~Varlet, D.~Bischoff, P.~Simonet, K.~Watanabe, T.~Taniguchi, T.~Ihn, K.~Ensslin, M.~Mucha-Kruczynski, and V.I.~Fal'ko, arXiv:1403.3244 (unpublished). 

\bibitem{Ben08}
C.W.J.~Beenakker, Rev.\ Mod.\ Phys.\ \textbf{80}, 1337 (2008).

\bibitem{gapfoo}
We further note that the result of Ref.\ \cite{May11}, reporting the conductivity quite close to $\sigma_\star$, suggests the influence of electron-electron interactions on charge transport in large ballistic bilayer samples may be negligible. 

\end{thebibliography}
\end{document}